\documentclass[twocolumn,10pt]{asme2e}

\usepackage{graphicx} 
\usepackage{comment}
\usepackage{amsmath}
\usepackage{wrapfig}
\usepackage{float}
\usepackage{url}
\usepackage{amssymb}
\usepackage[table,xcdraw]{xcolor}

\title{Convolutional Neural Network Model for Diabetic Retinopathy Feature Extraction and Classification}

\author{Sharan Subramanian
    \affiliation{
        s21sharan@gmail.com\\
	Merrill F. West High School\\
	Tracy, California 95376
    }	
}

\author{Leilani H. Gilpin\\
        lgilpin@ucsc.edu\\
        Computer Science and Engineering at UC Santa Cruz\\
        Santa Cruz, California 95064
}

\begin{document}
\maketitle    

\begin{abstract}
{\it The application of Artificial Intelligence in the medical market brings up increasing concerns but aids in more timely diagnosis of silent progressing diseases like Diabetic Retinopathy. In order to diagnose Diabetic Retinopathy (DR), ophthalmologists use color fundus images, or pictures of the back of the retina, to identify small distinct features through a difficult and time-consuming process. Our work creates a novel  CNN model and identifies the severity of DR through fundus image input. We classified 4 known DR features, including micro-aneurysms, cotton wools, exudates, and hemorrhages, through convolutional layers and were able to provide an accurate diagnostic without additional user input. The proposed model is more interpretable and robust to overfitting. We present initial results with a sensitivity of 97\% and an accuracy of 71\%.  Our contribution is an interpretable model with similar accuracy to more complex models.  With that, our model advances the field of DR detection and proves to be a key step towards AI-focused medical diagnosis. 
}
\end{abstract}

\section{Introduction}

Medical diagnosis serves as the foundational step in patient care, determining the pathway for treatment and intervention. According to Dr. Edmund Arthur of the University of Alabama at Birmingham School of Optometry, the accuracy and timeliness of diagnoses are essential, as they directly influence the patient’s prognosis and quality of life. A delay in identifying a condition can cause health issues, increase costs, and lower chances of full recovery. Recognizing a disease in its nascent stages often provides a broader array of treatment options and increases the likelihood of a favorable outcome. Diabetic Retinopathy (DR) exemplifies conditions where early detection can make a profound difference. DR, a complication resulting from diabetes, threatens the retina’s blood vessels, jeopardizing the crucial light-sensitive layer at the back of the eye. Despite the potential to prevent up to 98\% of vision loss with timely intervention, its silent progression often eludes timely detection, resulting in irreversible vision impairments.\par

\begin{figure*}
    \centering
    \includegraphics{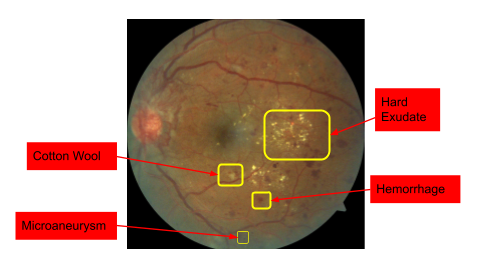}
    \caption{Different features of a CFP show the presence of DR. This Fundus Image was pulled from the Messidor-2 dataset and had a classified severity of 6 (PDR). These are the 4 features that were extracted and processed by our machine-learning model.}
    \label{fig:1}
\end{figure*}

Artificial Intelligence (AI) has emerged as a solution for more effective diagnostic tools. Within AI, there is a subset of complex models called Convolutional Neural Networks (CNNs).  However, CNNs applications for DR therapy have been of high complexity, requiring a high amount of computational resources. The most widely used approach for DR diagnosis consists of a dilated eye exam administered by an ophthalmologist or optometrist. However, deep learning models perform with the same accuracy as medical professionals at processing and analyzing fundus images\footnote{Charters, L. (2023). Finding a place for AI, machine learning in retinal imaging. \url{https://www.ophthalmologytimes.com/view/finding-a-place-for-ai-machine-learning-in-retinal-imaging
}}, which depict the retina in detail. CNNs may also be able to diagnose other conditions, including cataracts, glaucoma, and even illnesses outside the retina. In Figure~\ref{fig:1}, there are four main features from a fundus image showing positive signs of DR. By processing these images, CNNs can identify early indicators of DR, illuminating proactive medical interventions aided by AI.
In this paper, we present the different models currently available for DR detection and their limitations and challenges.  We review the technical methods used, including an in-depth explanation of the CNN architecture, our new methodology and model, and an open-source GitHub repository for public access\footnote{Github repository with the code for the proposed CNN model. This repository also has a detailed readme file on how to set up and use the model. \url{https://github.com/s21sharan/CNN_DR_Detection}}. The training process and post-training techniques used to improve accuracy are thoroughly examined in the Explanation of CNN Layers and Methods sections, respectively. In addition to the methodology, we compare pre-trained models with the proposed ADL-CNN model and examine the benefits and disadvantages of each. Furthermore, we explore the potential implications of CNN advancements in healthcare, suggest areas for further research, and discuss the impact these innovations can have on disease detection and management.

\section{Literature Review}
We provide a brief history of DR detection and how its diagnosis evolved. In addition to going over the various pre-existing studies on CNN models for DR detection, we also highlight each model’s and method’s limitations. Then, we will highlight what our model brings that is different from other studies and the advantages our model entails.
\subsection{Historical overview of DR detection}
In the early stages of DR diagnosis, its approach was based on a patient’s symptomatic presentation and rudimentary examination techniques \cite{Wolfensberger2001-ns}. Eduard Jaeger and Albert von Graefe, in the 1850s described diabetic macular changes: a historical review suggests the evolution of our understanding\cite{Wolfensberger2001-ns}. It took more than a century for a consensus to emerge regarding the link between cystoid macular edema, which is caused by the accumulation of fluids in the macula, and diabetes mellitus (DM), where the body's immune system attacks and destroys insulin-producing cells in the pancreas. Recognizing the connection between cystoid macular edema and diabetes mellitus today influences diagnostic methods for patients with visual symptoms.\par
The first line of diagnosis typically involved direct ophthalmoscopy, a method discussed in the historical evolution of DR diagnosis\cite{Matuszewski2017-tq}. Digital fundus photography reshaped DR screening in the late 20th and early 21st century. Comparisons between singly, two-field, and three-field 45-degree Color Fundus Photographs (CFP) showed varying sensitivities and specificities. These advancements have modernized DR diagnosis, with varying sensitivities and specificities observed across different Color Fundus Photograph methods.\par
With more advances in integrated AI and DR screening, the FDA approved an autonomous AI device that used 45-degree digital CFP for DR detection called the EyeArt AI system. On a side note, confocal scanning ophthalmoscopy emerged as a technique that provided ultra-widefield imaging.\par
With these advancements, other challenges lie ahead on this path of innovation and integration. The intricacies of integrating various technologies seamlessly, ensuring universal accessibility to these advanced diagnostic tools, and refining AI algorithms for increased accuracy are among the ongoing challenges. While previous models may have offered solutions to specific challenges, a collaborative effort across medical, technological, and regulatory domains is essential for realizing the full potential of these novel diagnostic approaches.
\subsection{Specific Studies on DR Detection using CNNs}
In a study by Yasashvini R. et al., researchers attempted to classify DR using both CNN and hybrid deep CNNs\cite{Yasashvini2022-bw}. Their methodology combined the strengths of CNN architectures with a hybrid deep CNN via a mixture of different classifiers. They hypothesized that by combining the power of both these mixed classifiers, the accuracy of DR would increase, better accounting for the variety of features and abnormalities present in DR-affected CFPs. Their results were successful in achieving accuracy rates close to 100\%.\par
\begin{figure}
    \centering
    \includegraphics[width=\linewidth]{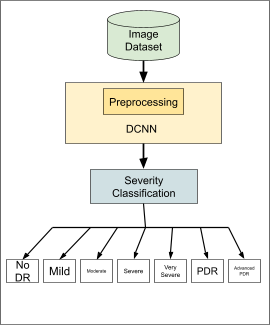}
    \caption{CNN architecture for a DR detection model. This model finds the severity of DR through 7 different classifications of DR (No DR, Mild DR, Moderate DR, Severe DR, Very Severe DR, PDR, and Advanced PDR).}
    \label{fig:2}
\end{figure}

Following the onset of high precision in DR-prediction models, Pratt et al. introduced the task of predicting severity in DR diagnosis\cite{Pratt2016-xe}. Their multi-dimensional data supports severity prediction in CFPs, from retinal features such as micro-aneurysms, exudates, and hemorrhages. Pratt and his team developed a CNN architecture that underwent data augmentation to achieve this dual objective. In Figure \ref{fig:2}, we can see the architecture that Pratt et al. used. Their CNN model classified the fundus images between the seven different severities with a sensitivity of 95\% and an accuracy of 75\% on the validation images.  In our work, we augment their architecture’s weights and layers.\par

Albahli \& Yar's research emphasized the customizability of CNNs in the medical domain\cite{Albahli2022-lf}. Albahli \& Yar designed three deep-learning models to detect the severity of diabetic retinopathy from retina images and determine its potential progression to macular edema. Their main challenge was the limited dataset they used: the Indian Diabetic Retinopathy Image Dataset (IDRiD). To counteract this, they designed features such as Brightness, Color, and Contrast enhancement, Color Jitters, and Contrast Limited Adaptive Histogram Equalization to generate a broader range of images. Their study used pre-trained models like ResNet50, VGG16, and VGG19 and customized them to DR detection. Their results, after validation, showcased the potential of custom CNNs, with each model yielding reasonable outputs.\par
On the other hand, the study by Shu-I Pao et al. used entropy images to increase DR detection performance\cite{Pao2020-wb}. In this context, the entropy image is an abstract representation computed using the green component of the CFP. By preprocessing these entropy images using unsharp masking, the researchers were able to enhance the clarity and definition of the features. The outcome was a bi-channel CNN, which effectively utilized the features from the gray level and the green component's entropy images. This approach, while complex, provided a comprehensive manner for improving the detection performance of severe DR cases.\par
In summary, these studies show the immense potential and versatility of CNNs in DR detection. Whether it is the combined powers of CNNs and hybrid networks, the emphasis on severity classification, the adaptability of custom models, or the complex manipulation of entropy images, the domain of DR detection is witnessing fast progress and growth.

\subsection{Limitations of Current Models}
One explicit limitation emerges from the study by Yasashvini et al. While integrating standard CNNs with hybrid deep convolutional networks might enhance precision, it inadvertently adds to the model's complexity. Such sophisticated models often require vast computational resources, both in terms of training and inference. Furthermore, the more complex a model is, the higher the likelihood of overfitting, especially when the available dataset is not large enough. Our proposed model is less complex and has repeatedly been tested on overfitting curves. Overfit models might perform exceedingly well on training data but falter when faced with unseen data, making them unreliable for real-world applications. Testing the model with multiple data sets to ensure uniform reliability in the real world is important.\par
In the research led by Pratt et al., the results were primarily based on one specific dataset: the Kaggle dataset. The variability and uniqueness of retinal images suggest that a model trained on one dataset might not generalize well to images from different sources, patients of varied demographics, or images captured using different equipment. This bias-related limitation ties into the data scarcity in DR and fundus imaging.\par
Albahli's research, which introduced custom CNN models, highlighted another significant limitation related to data scarcity. The requirement to employ image generation techniques like BCC enhancing, CJ, and CLAHE to expand their dataset indicates the prevalent challenge of limited datasets in the field. Having a limited dataset not only hampers the depth of training but also raises concerns about the model's ability to generalize to a broader range of images. In our model, we have trained and tested it with a more comprehensive dataset and have seen repeatable and reliable results with additional data sets. Moreover, artificially augmenting data can sometimes introduce new artifacts or fail to replicate the subtle nuances of natural images, potentially leading to models that might misinterpret or overlook certain features. Another question arises by introducing many data augmentation techniques: whether pre-processing techniques mask significant data features.\par
Shu-I Pao and his team's research brings forth the intricacies of preprocessing. While the emphasis on entropy images derived from the green component of fundus photographs and preprocessing techniques like unsharp masking is innovative, it also adds layers of processing that might not be feasible in real-world, time-sensitive scenarios. Such intricate preprocessing might also introduce biases or inadvertently filter out certain relevant features, limiting the model's diagnostic capabilities. Our model sticks to basic preprocessing to resize the image and extract key features without modifying or filtering the actual image itself.
Lastly, a collective limitation observed from all these studies is the gap between theoretical performance and real-world implementation. While these models showcase impressive metrics in controlled environments, challenges such as computational constraints, integration with existing healthcare systems, interpretability of results, and ensuring consistent performance across diverse patient populations remain. Addressing these practical limitations is paramount for the successful and widespread adoption of CNNs in DR diagnostics.\par

\subsection{AI Trust in Healthcare Environment}
The area of healthcare, particularly in recent years, has observed a large fusion of artificial intelligence (AI) applications, covering a broad spectrum ranging from diagnostics and clinical decision-making to digital public health. AI's potential to enhance clinical outcomes and efficiency is acknowledged, but the acceptance of this technology remains contingent on multiple factors.\par
During the  COVID-19 pandemic, Chalutz Ben-Gal's study investigated the acceptance of AI in primary care\cite{Chalutz_Ben-Gal2022-yj}. It was found that while many patients showcase a resistance towards AI, several factors can potentially tilt the balance in favor of AI. Explicit mentions of AI's superior accuracy, positive nudges from primary care physicians, and assurances of personalized patient listening experiences could enhance AI acceptance. Notably, Robertson et al. found that specific demographics, including Black respondents and older individuals, were less inclined to choose AI, highlighting the consideration of demographic-specific concerns in AI adoption\cite{Robertson2023-pm}.\par
Diving deeper into the diagnostic field, the potential of AI to augment diagnostic precision is immense. However, as noted by Robertson et al., patients' trust in AI may waver based on disease severity and their perceived personalization of the AI system.
Lastly, considering the viewpoint of healthcare professionals, Lambert et al. presented an integrative review highlighting the barriers to AI acceptance in the hospital setting\cite{Lambert2023-te}. The fear of losing professional autonomy in integrating AI into existing clinical workflows was a recurring concern. Conversely, training geared towards AI utilization emerged as a positive catalyst and gained acceptance among countless professionals. The study further emphasized the significance of involving end-users during the early stages of AI development, suggesting a collaborative approach to ensure that the technology meets real-world needs and challenges.\par
In conclusion, while the potential benefits of AI in healthcare are vast, its acceptance remains an issue, necessitating a comprehensive understanding of both patients' and practitioners' perspectives coupled with tailored approaches to address their specific concerns.

\section{Understanding Convolutional Neural Networks}
Convolutional Neural Networks (CNNs) are pivotal in machine learning and artificial intelligence, especially for image processing and classification tasks. Inspired by biological neural networks, CNNs emulate human perception by recognizing patterns from image pixels, enabling classification, identification, and visual data reconstruction \cite{Shin2016-dq}. In the realm of healthcare, particularly radiology, they exhibit considerable potential in enhancing conventional diagnostic methodologies, resulting in improved speed, reliability, and occasionally superior accuracy\cite{Yamashita2018-mq}.\par
\begin{figure*}
    \centering
    \includegraphics[width=\textwidth]{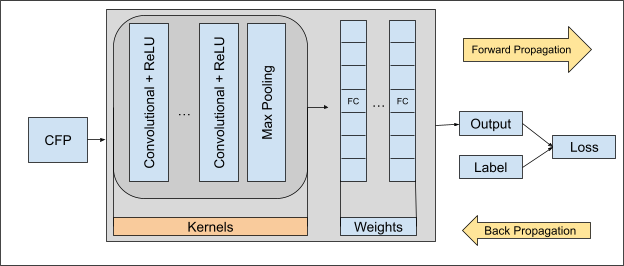}
    \caption{This flowchart highlights the CNN architecture that is used in our proposed model. It has different kernels and depth layers that take input images and output the classification. The input layer is where the color fundus image is fed into the neural network. Then the convolutional layers apply a set of filters to the input image, detecting specific features like edges, textures, and shapes that are tied into the DR signs, as highlighted in Figure 1. Then, the ReLU activation function is applied element-wise, introducing nonlinearity. The weights represent the parameters that the network learns during training, and the final output is the severity classification of DR based on features extracted from the input image by preceding layers.}
    \label{fig:3}
\end{figure*}

Fundamentally, neural networks comprise interconnected nodes or "neurons" that execute elementary mathematical operations. These networks are constructed to model intricate data relationships. As depicted in Figure \ref{fig:3}, CNNs encompass input layers for data reception, hidden layers for data processing, and output layers for prediction/decision-making. Unlike conventional neural networks employing full interconnections between layers, CNNs employ sparse connections to focus on local features, rendering them highly efficient in capturing spatial hierarchies. Consequently, CNNs are exceptionally well-suited for tasks such as image recognition, where the extraction of local features such as edges, corners, and textures offers valuable information.\par
CNNs belong to a category of neural networks known as Deep Learning models\cite{Alzubaidi2021-xe}. As the name suggests, these models are characterized by their depth or the number of layers data must pass hierarchically. Deep Learning, a subfield of machine learning, offers significant advantages over traditional machine learning methods. The depth of the network allows for more complex features to be learned, thereby increasing the model's ability to understand intricate patterns in the data, especially in medical imaging tasks\cite{Yala2019-ad}.

\subsection{Breaking Down the Architecture: Layers, Filters, Feature Extraction, etc.}
The convolutional layer is the core aspect of a CNN. This layer performs the convolution operation on the input, passing the result to the next layer. These operations focus on local regions of the input, preserving the relationship between pixels by learning image features\cite{Taye2023-lg}. For instance, in DR detection, this localized focus enables CNNs to discern pixel-level details, capturing critical retinal features and enhancing the model's feature-learning capability.\par
After convolution, the layers often employ an activation function, the most common being the Rectified Linear Unit (ReLU). ReLU, in the kernels section of Figure 3, introduces non-linearity into the system, which allows the model to learn from the error and make adjustments, improving its performance during the training process.\par
Following the convolutional and activation layers, CNNs usually contain pooling layers to reduce dimensionality. The pooling operation condenses the feature map obtained from the previous layer, usually by taking the maximum (as in Equation \ref{eq:1}) or average value (as in Equation \ref{eq:2}) from a group of values in the feature map\footnote{Nanos, G., \& Nanos, G. (2023). Neural Networks: Pooling layers, Baeldung on Computer Science. Baeldung on Computer Science. \url{https://www.baeldung.com/cs/neural-networks-pooling-layers}}. This pooling process also results in reduced size and computational complexity of the resultant feature map.\par
\begin{equation} 
    {MaxPooling}(X)_{i,j,k} = \max_{m,n} X_{i \cdot s_x + m, j \cdot s_y + n, k} 
    \label{eq:1}
\end{equation}
\begin{equation} 
    {AvgPooling}(X)_{i,j,k} = \frac{1}{f_x \cdot f_y}\sum_{m,n} X_{i \cdot s_x + m, j \cdot s_y + n, k}
    \label{eq:2}
\end{equation}
One of the significant operations in CNNs involves filter or kernel feature extraction, as illustrated in Figure \ref{fig:3}. Filters, which are small and adjustable parameters, traverse the input image, generating feature maps. For example, one filter may focus on edge detection, while another specializes in identifying specific colors. The concept of "stride" defines how filters move across the input image, determining their step size. These adjustable parameters collectively provide spatial control over the convolution operation, enabling precise adjustments and customization.\par
Towards the end of the network, fully connected layers, also known as dense layers, utilize the high-level features acquired from preceding layers to classify the image into distinct categories. These layers share a structural resemblance to a standard multi-layer perceptron neural network.

\subsection{Explanation of CNN Layers, with a Specific Focus on Image Classification Tasks}
Forward propagation in CNNs entails the sequential transmission of the input image through multiple layers. As the image traverses from one layer to the next, it gives rise to feature maps characterized by reduced dimensionality, which are subsequently employed in the final layers for classification purposes\cite{Peng2021-fp}. Similar to other neural networks, CNNs use backpropagation for training. This algorithm evaluates the disparity between the output produced by forward propagation and the expected outcome, subsequently modifying the filter weights, as illustrated in Figure \ref{fig:3}. This weight adjustment process is often facilitated by Stochastic Gradient Descent (SGD) optimizers.\par
\begin{enumerate}
    \item Initialization: Initialize the weights (\(w_1\), \(w_2\)) and biases (\(b_1\),\(b_2\)) with random values.
    \item Forward Pass: Pass a training example (\(x_train\)) through the network to compute the predicted output (\(y\))
    \item Compute Loss: Calculate the loss, as in equation \ref{eq:3}, using predicted output and the true target
        \begin{equation}
            L=\frac{1}{2}(y-y_{true})^2
            \label{eq:3}
        \end{equation}
    \item Backpropagation: Compute the gradients of the loss with respect to the weights and biases using the chain rule (calculus)
    \item Update Weight: Adjust the weights using the gradients and small learning rate according to SGD update rule\footnote{Ruder, S. (2020). An overview of gradient descent optimization algorithms. ruder.io. \url{https://www.ruder.io/optimizing-gradient-descent/}}, as in Equation \ref{eq:4} and \ref{eq:5}
        \begin{equation}
            w_i = w_i - \alpha (\frac{\vartheta L}{\vartheta w_i})
            \label{eq:4}
        \end{equation}
        \begin{equation}
            b_i = b_i - \alpha (\frac{\vartheta L}{\vartheta b_i})
            \label{eq:5}
        \end{equation}
    \item Repeat: Repeat steps 2-5 for specified number of epochs until convergence
\end{enumerate}
Pre-trained models are applied in various scenarios, particularly when working with limited training datasets. This approach applies pre-trained CNNs, with subsequent fine-tuning of their final layers for adaptation to new tasks. This reduces computational resource demands instead of using significant compute to train a CNN from scratch. CNNs utilize batch processing techniques, which enhance training efficiency by enabling faster and more parallel computation. Furthermore, cross-entropy loss functions are used to quantify the disparity between predicted outputs and actual labels, with the primary training objective being the minimization of this loss function. Without the cross-entropy loss function, CNNs risk overfitting, introducing biases, and failing to generalize effectively on unseen data. Similarly, a CNN's performance can be influenced by hyperparameters like learning rate, filter size, filter count, and layer count. Consequently, hyperparameter optimization is frequently conducted to refine these settings.\par
Beyond image classification, CNNs find applications in complex tasks such as video analysis, natural language processing, and autonomous driving. In the healthcare domain, their image recognition capabilities are invaluable for the detection of anomalies in X-rays, MRIs, CT scans, and CFPs, as elaborated in this paper. As CNNs and their associated technologies continue to evolve, they hold immense promise for healthcare applications, ranging from early disease classification for conditions like cataracts and cancers to real-time patient monitoring.

\section{Methodology}
\subsection{Data Sets Used}
\begin{table}[]
    \resizebox{\linewidth}{!}{%
    \begin{tabular}{|
        >{\columncolor[HTML]{FFFFFF}}l |
        >{\columncolor[HTML]{FFFFFF}}l |
        >{\columncolor[HTML]{FFFFFF}}l |
        >{\columncolor[HTML]{FFFFFF}}l |l}
        \cline{1-4}
        \cellcolor[HTML]{9B9B9B}Dataset & \cellcolor[HTML]{9B9B9B}Year Released & \cellcolor[HTML]{9B9B9B}Image Count & \cellcolor[HTML]{9B9B9B}Field of Vision (FOV) &  \\ \cline{1-4}
        Kaggle     & 2015 & 88k   & 50°           &  \\
        APTOS      & 2019 & 13k   & Not specified &  \\
        DDR        & 2019 & 13.6k & 45°           &  \\
        Messidor-2 & 2010 & 1748  & 45°           &  \\
        DeepDRid   & 2019 & 2256  & Not specified &  \\ \cline{1-4}
    \end{tabular}%
    }
    \label{tbl:1}
    \caption{The different publicly available CFP datasets for classifying DR.}
\end{table}
We incorporated multiple data sets to develop, train, and validate our model, with a primary focus on the Messidor-2 dataset. Table 1 gives a snapshot of the datasets we utilized.
The primary dataset which our model was built from is the Messidor-2 dataset. Between October 2009 and September 2010, diabetic patients were imaged using a Topcon TRC NW6 non-mydriatic fundus camera with a 45° field of view, resulting in 345 total DR examinations. Only macula-centered images were considered for this dataset. This was then combined with the Messidor-original database, which has a total of 529 DR examinations, to create the Messidor-2 Data set.\par
Initially, we used a 70-30 train-test split with the Messidor-2 dataset taking inspiration from other studies in the field\cite{Mutawa2023-ek}. However, we adjusted this split to 80-20 to minimize training and validation losses.\par
\begin{figure}
    \centering
    \includegraphics[width=\linewidth]{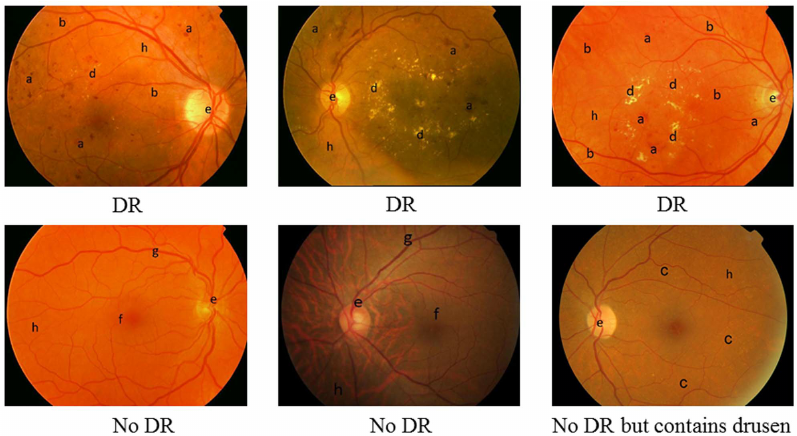}
    \caption{This figure has three images of CFPs that contain DR (in the top row) and three images of CFPs that do not contain DR (in the bottom row). The annotated features are as follows: a. Hemorrhage; b. MA; c. Drusen; d. Exudates; e. Optic disc; f. Fovea; g. Blood vessel; h. Background doi: https://doi.org/10.1371/journal.pone.0066730.g001}
    \label{fig:4}
\end{figure}

Our secondary data sets include the Kaggle dataset, APTOS, DDR, and DeepDRid. 
\begin{enumerate}
    \item Kaggle: This dataset was involved in one of the Kaggle Challenges relating to medical imaging and CNNs. This dataset covers a diverse range of sources and conditions, with over 88,000 images.
    \item APTOS (Asia Pacific Tele-Opthalmology Society): This society concentrates on eye-related conditions, including DR. This dataset includes patients of Asian descent and is very comprehensive, with over 13k images.
    \item DDR: This dataset is sourced from 147 hospitals spread across 243 provinces in China. These images are classified based on the severity of diabetic retinopathy into 5 categories.
    \item DeepDRid: This dataset originated from the 2nd Diabetic Retinopathy: Segmentation and Grading Challenge in collaboration with ISBI in 2018. This dataset focuses on three pivotal tasks: dual-view disease grading, image quality estimation, and transfer learning.
\end{enumerate}
All these datasets were classified based on the severity of DR present in the images. The five severities were none, mild, moderate, severe, and proliferative DR. The images of poor data quality were excluded from the datasets, and a manual review took place before putting these images into our model. In Figure \ref{fig:4}, we can see examples of CFPs that passed this review and made it onto the dataset. There are significant differences between these CFPs and their features, including the hard exudate, cotton wool, and hemorrhage, as annotated in Figure \ref{fig:1}.

\subsection{Methods}
Our approach is split into two primary stages: the training/validation phase and the testing phase.\par
First, Data was aggregated from various sources like hospitals and specialized data repositories during the initial phase. We used the Messidor-2 dataset as our baseline data. We removed any 25 low-quality images from the consolidated dataset through an ancillary intelligent model and organized them using a 70-30 split\footnote{Our updated dataset is attached to the github link in Footnote 2.}. We split each subcategory, or severity, using Sklearn’s train-test split function. \par
During the pre-processing stage, we used two operations to reduce noise and remove gaps in the background and foreground of the CFP. The first procedure is known as Erosion. Erosion is used to eliminate or spike the edge of the area and is represented by Equation \ref{eq:6}\cite{Sarki2021-mq}.
\begin{equation}
    A\ominus B={p|B_p\subseteqq A}
    \label{eq:6}
\end{equation}
Then we also applied dilation, which is used to broaden the rim of the background or foreground image configuration. This procedure helps us fill the gap created through erosion and is defined by  equation \ref{eq:7}\cite{Sarki2021-mq}:
\begin{equation}
    A\oplus B={x|B_x\cap X\neq 0}
    \label{eq:7}
\end{equation}

Where $$\ominus$$ denote the dilation; $$\oplus$$ denote the erosion; \(A\) = Structuring element and \(B\) = the erosion of the dilation of that set. These equations are from the \textit{Tyler Coye Algorithm}.\par
In the CNN model, we used convolutional operations to extract features. We could convert the images into feature activation maps using diverse filter sizes. Then, using the Rectified Linear Unit (ReLU) activation function, we further enhanced this model with non-linearity. Following this, we employed a pooling layer, an instrumental part in compressing the activation map’s dimensions without significant information loss. We then vectored the activation maps, processed through several fully connected layers, resulting in a severity rating based on the DR-positive or DR-negative classification.

In the model’s second stage, we tested the data using our test-split of the Messidor-2 data set and other secondary data sets, as highlighted in section \ref{Data Sets Used}. To prevent our model from overfitting, we implemented the early stopping mechanism, which was used to reduce training time from 8 hours 36 mins to 4 hours and 55 mins. This technique monitors our model’s performance on the validation set and halts training once the performance starts deteriorating. This ensures that the model does not overfit. We also implemented a drop-out mechanism, which allowed randomly selected neurons to be ignored at each iteration. This ensured that our neural network remained accurate and generalized well to new-unseen data.\par
We also used the Keras auto-hyperparameter tuning module to tune crucial parameters of our models. This algorithm uses a mathematical approach to parameter tuning; this tweaks the learning rate, batch size, epochs, and dropout rate to ensure the lowest validation losses and highest accuracy. We were able to increase our validation accuracy from 55\% to 70\% using this hyperparameter tuner.
\section{Results}
In our initial study, the DR model was trained, validated, and tested on the Messidor data set. With 1748 images split into five severity categories, we had about 300 for each severity. These 500 images were then split in an 80-20 ratio for the test-train split.\par
\begin{figure*}
    \centering
    \includegraphics[width=\textwidth]{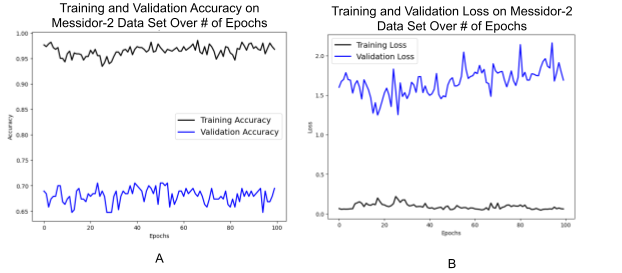}
    \caption{Figure \ref{fig:6}a shows the Accuracy curves for our model as tested on the Messidor-2 Data set. Figure \ref{fig:6}b shows the Loss curves for our model as tested on the Messidor-2 Data set.}
    \label{fig:6}
\end{figure*}
As seen in Figure \ref{fig:6}a, our model’s performance throughout the training process is consistent. Averaging at approximately 97\% over 100 epochs, this model can recognize and predict patterns within the training dataset. The validation accuracy remained at  70\% throughout the 100 epochs. Though slightly lower than the training data, this validation accuracy aligns with the understanding that generalizing our model to unseen data will reasonably lower the accuracy.\par
The training loss, as seen in Figure 5b, remained under 0.3 across all epochs. On the other hand, the validation loss was higher at about 1.5-1.8. As shown in Table \ref{tbl:2}, the training and validation accuracy also stabilized around the seventh epoch, hinting that our accuracies were not just a fluke.\par
Additionally, this model was reset and repeatedly tested, bringing stable and similar accuracy values every time.

\section{Discussion}
Our model demonstrates the effectiveness of CNNs in DR detection, with a streamlined architecture. In comparison to Yasashvini R. et al.'s hybrid deep CNNs, our model strikes a balance between complexity and efficacy. Unlike Albahli et al.'s research, which relied on multiple image enhancement techniques due to data scarcity, our model was trained on a more comprehensive dataset and validated on additional datasets.\par
In contrast to techniques like entropy-based image processing highlighted by Shu-I Pao et al., our model aims for direct, effective, and reproducible DR classification, maintaining competitive validation accuracy.\par
\begin{table}[H]
    \resizebox{\columnwidth}{!}{%
    \begin{tabular}{|l|l|l|l|l|}
        \hline
            \rowcolor[HTML]{B7B7B7} 
            Epochs                    & Training Loss                    & Training Accuracy                & Validation Loss                  & Validation Accuracy \\ \hline
            \cellcolor[HTML]{FFFFFF}0 & \cellcolor[HTML]{FFFFFF}0.075937 & \cellcolor[HTML]{FFFFFF}0.971781 & \cellcolor[HTML]{FFFFFF}1.577781 & 0.700000            \\
            \cellcolor[HTML]{FFFFFF}1 & \cellcolor[HTML]{FFFFFF}0.076237 & \cellcolor[HTML]{FFFFFF}0.964727 & \cellcolor[HTML]{FFFFFF}1.689509 & 0.694737            \\
            \cellcolor[HTML]{FFFFFF}2 & \cellcolor[HTML]{FFFFFF}0.083331 & \cellcolor[HTML]{FFFFFF}0.970018 & \cellcolor[HTML]{FFFFFF}1.463975 & 0.710526            \\
            \cellcolor[HTML]{FFFFFF}3 & \cellcolor[HTML]{FFFFFF}0.076263 & \cellcolor[HTML]{FFFFFF}0.973545 & \cellcolor[HTML]{FFFFFF}1.697605 & 0.700000            \\
            \cellcolor[HTML]{FFFFFF}4 & \cellcolor[HTML]{FFFFFF}0.96008  & \cellcolor[HTML]{FFFFFF}0.962963 & \cellcolor[HTML]{FFFFFF}1.778189 & 0.663158            \\
            5                         & 0.073751                         & 0.966490                         & 1.582166                         & 0.693684            \\
            6                         & 0.072941                         & 0.964727                         & 1.623718                         & 0.784211            \\
            7                         & 0.076225                         & 0.966490                         & 1.609345                         & 0.700000            \\ 
        \hline
    \end{tabular}%
    }
    \label{tbl:2}
    \caption{Training and Validation Results for Proposed CNN model}
\end{table}
A 20\% delta between training and validation metrics indicates room for improvement. Future work includes implementing regularization techniques like L1 or L2 regularization to address this disparity. Additionally, inspired by Su-I Pao et al.'s use of entropy images, we may explore further feature engineering techniques to extract informative features from retinal images.\par
In the evolving field of DR detection, our model offers promise and scalability. Future iterations, along with advancements such as regularization techniques, have the potential to improve accuracy. Our approach's simplicity enables integration with diverse datasets and applicability across demographics and equipment sources.

\section{Conclusion}
In our study on DR detection using CNNs, we achieved enhanced efficiency and accuracy in categorizing DR severity levels from retinal images. Our approach, which incorporated preprocessing techniques like erosion and dilation alongside convolutional operations, consistently yielded a 97\% training accuracy over 100 epochs. Impressively, our validation accuracy reached 71\%, marking significant progress towards automated medical diagnosis in DR.\par
Comparative analysis with other studies underscores the effectiveness of our approach, striking a balance between model understanding and detection efficacy. The model is adaptable across diverse datasets, requiring minimal image enhancement due to the comprehensive Messidor-2 dataset used. However, the gap between training and validation metrics indicates potential for model enhancement, including advanced regularization techniques and feature engineering to improve accuracy and mitigate overfitting.\par
Future improvements may explore more advanced preprocessing techniques and incorporate regularization methods to bridge the gap between training and validation loss metrics. Collaborating with healthcare institutions to obtain diverse datasets could further enhance the model's adaptability across various clinical scenarios.\par
Our consistent results and the interpretability of our approach underscore the reliability of CNNs in DR detection. Our work raises questions about how healthcare professionals will adapt to AI technologies in the medical field and the potential for more optimal Neural Networks in medical classification tasks. As the medical community embraces the intersection of healthcare and AI, the advancements presented in this and other studies contribute to a safer and more clinically advanced society.

\nocite{*}
\bibliographystyle{style}
\bibliography{asme2e}

\end{document}